# Role of Electronic Structure in the Morphotropic Phase Boundary of $Tb_xDy_{1-x}Co_2$ Studied by First-principles Calculation

Dongyan Zhang[1,2,a], Sen Yang[1,b], Pangpang Wang[3], Yu Wang[1], Jieqiong Wang[1], Xiaoping Song[1]

[1]State Key laboratory for Mechanical Behavior of Materials and MOE Key Laboratory for Nonequilibrium synthesis and Modulation of Condensed Matter, Xi'an Jiaotong University, Xi'an 710049, People's Republic of China

[2]Graduate School of Advanced Technology and Science, the University of Tokushima, Tokushima 770-8506, Japan

[3]Institute for Materials Chemistry and Engineering, Kyushu University, Fukuoka 812-8581, Japan

**Abstract**

Physically parallel to ferroelectric morphotropic phase boundary, a phase boundary separating two ferromagnetic phase of different crystallographic symmetries was found in $Tb_xDy_{1-x}Co_2$. High-resolution synchrotron XRD has been carried out to offer experimental evidence for $Tb_xDy_{1-x}Co_2$. It has been proved that $Tb_xDy_{1-x}Co_2$ (0.6<x<0.7)

Author to whom correspondence should be addressed.
a) E-mail:zhang@tokushima-u.ac.jp
b) E-mail: yangsen@mail.xjtu.edu.cn

19  is a morphotropic phase boundary and that the crystal structures of tetragonal (x<0.6)

20  and rhombohedral (x>0.7) phase is distorted from a Laves Phase. Here, a first principles

21  calculation provides a theoretical explanation on the origin of MBP in $Tb_xDy_{1-x}Co_2$ and

22  is also provided for the question of why MPB occurs in $Tb_xDy_{1-x}Co_2$ alloys.

23

24

RM$_2$ (R: rare-earth element, M: transition metal) with the cubic Laves phase structure have attracted great attention for their variety of magnetic behavior[1-3]. All of them possess very similar lattice parameters[4] and exhibit a magnetic instability of the 3d-subsystem due to the occupation of $d$ and $f$ orbitals[5]. Investigation on electronic structure of RM$_2$ would improve our understanding on mechanism of their magnetic behavior. With the more results provided by experimental progress, the theoretical analysis made it possible to understand systematically the influence of the electronic structure on the magnetic properties[6]. A lot of researches[7-10] have been contributed on the theoretical and experimental investigation of RFe$_2$, RCo$_2$ and RNi$_2$. Especially for Tb$_x$Dy$_{1-x}$Co$_2$, a debate on the phase transition, if is the spin-reorientation transition also a border separating different crystal symmetries, has been continued for recent decades. The progress in this issue provides a deep insight into the nature of a fundamental concept in magnetism and provides a chance to study the transitions associated with electronic structure. Specially, the magnetic properties of cobalt are strongly depended on the rare-earth element .It offers an opportunity to investigate the occupation of $d$ and $f$ orbitals and hybridization between them. As early as 1970s', Bloch[11] have pointed out that the cobalt moment is not intrinsic but induced in the d band by the exchange field due to the rare-earth, the behavior of transition in RCo$_2$ have related the density of states at the Fermi Level. Recently, Yang[12-14] has proved that a similar MPB situation exist in the ferromagnetic system as same as in the ferroelectric system, which have rectified the long standing mistake that the magnetic transition in $Tb_xDy_{1-x}Co_2$ alloy is merely a reorientation of magnetization direction and there is no change in crystal symmetry. This promises to change our perception of how we think about the magnetic theory. Yang employed the high-resolution synchrotron XRD at the BL15XU NIMS beat line of Spring-8 to experimentally confirm the transition in $Tb_xDy_{1-x}Co_2$ alloy. However, most of the theoretical explanations concerning the underlying mechanism of MPB have been phonological studies[15, 16]. In this letter, we investigated the electronic structure at the Fermi Level in $Tb_xDy_{1-x}Co_2$ and clarified the origin of different ferromagnetic states corresponding to different crystal symmetries in $Tb_xDy_{1-x}Co_2$ alloy by using first principles calculation. This work provides a theoretical explanation on the origin of MBP in $Tb_xDy_{1-x}Co_2$ and is also provided for the question of why MPB occurs in $Tb_xDy_{1-x}Co_2$ alloys.

The calculations have been performed on the basis of spin-polarized DFT[17, 18] with the *ab initio* total energy program CASTEP[19]. For the exchange-correlation functional, we chose a local density approximation in the scheme of Ceperley and Alder[20] parameterized by Perdew and Zunger[21] denoted hereby CA-PZ. The spin interpolation was used in this calculation. The self-consistent ground state of the system was determined using a band-by-band conjugate gradient technique to minimize the total energy of the system with respect to plane-wave coefficients. The electronic wave functions were obtained by the All Bands/EDFT scheme. A number of twelve valence electrons for each Dy atom ($4f^{10}6s^2$) and eleven valence electrons for each Tb atom ($4f^9 6s^2$) were taken into account. For each Co atom ($3d^7 4s^2$), a number of nine valence electrons were employed and the formal spin state with 4 minority spin electrons was used as initial. The remaining core electrons together with the nuclei were described by pseudo-potentials in the framework of the PAW method[22, 23]. The one-electron Kohn-Sham wavefunctions as well as the charge density were expanded in a plane-wave basis set. The initial crystal structure model of cubic $TB_xDy_{1-x}Co_2$ was build according to the previous research of the synchrotron XRD data[14]. Then, the Dy atoms in the supercell were gradually replaced and the cell was geometry optimized. In order to optimize the total energy of $TB_xDy_{1-x}Co_2$ with Tetragonal symmetry, the crystal constant c was only relaxed, and the other degrees of freedom were frozen. Similarly, in order to optimize the total energy of $Tb_xDy_{1-x}Co_2$ with Rhombohedral symmetry, the crystal constant *α,β* and *γ* were only relaxed, and the crystal constant a, b and c were fixed.

The lattice parameters for $DyCo_2$ and $TbCo_2$ at 0 K were extrapolated from the low

79　temperature X-ray analysis. A model of DyCo2 in Laves phase was built with the lattice

80　constant 7.12 A[24, 25], and TbCo2 in Laves phase with lattice constant 7.25 A[26]. Then, the

81　special symmetry employed on the model in Laves phase was canceled, and atoms in

82　fractional coordinate were constrained. The crystal structure model of $Tb_xDy_{1-x}Co_2$ with

83　tetragonal symmetry or rhombohedral symmetry was modified from the Laves phase.

84　Figure 1 shows the Laves phase configuration and the DOS for $DyCo_2$ with Laves phase.

85　The crystallographic cell consists of an eight $Tb_xDy_{1-x}Co_2$ formula unit. The total

86　energies of $Tb_xDy_{1-x}Co_2$ with Tetragonal symmetry were calculated with the

87　compressing or stretching the *c* axis of Laves phase, and the total energies of

88　$Tb_xDy_{1-x}Co_2$ with Rhombohedral symmetry were calculated with altering the crystal

89　constant *α,β* and *γ* of Laves phase, synchronously. The c-axis was distorted to calculate

90　the total energy of $Tb_xDy_{1-x}Co_2$ with AMD symmetry. Otherwise, the angle was

91　distorted to calculate the total energy of $Tb_xDy_{1-x}Co_2$ with R-3M symmetry. The total

92　energies of $Tb_xDy_{1-x}Co_2$ with Tetragonal symmetry and Rhombohedral symmetry were

93　shown in Figure 2. The horizontal coordinate indicates the distortion of *c*-axis. The

94　positive value reveals the compression, and the negative value reveals the stretch. The

95　ground state of the $Tb_xDy_{1-x}Co_2$ was determined by considering spin configurations. All

96　magnetic moments are aligned parallel in the ferromagnetic (FM) state for a

97  crystallographic cell, which is consistent with the FM nature of $Tb_xDy_{1-x}Co_2$ under the

98  Curie temperature.

99      Firstly, on the basis of atom configuration of Laves Phase, we investigated the

100 electronic structure of $DyCo_2$. Figure 1 shows the Laves phase configuration and the

101 DOS for $DyCo_2$. For one stoichiometric cell, twelve Co atoms locate at the vertex of a

102 truncated tetrahedron, and one Dy atom locates at the center. The inset diagram is the

103 schematic of spin electron configuration for Co atom in the truncated tetrahedron. In the

104 view of classical crystal field theory, due to the Co's occupation at the vertex of a

105 truncated tetrahedron rather than the central, the distortion of the crystal structure would

106 not affect the spin electronic configuration of Co. It implies that there would be no

107 structure phase transition in the case of Dy's substitute by Tb. It is significantly far from

108 the experimental observation and our results of total energy calculation. Prof. Kimura[27]

109 has developed an *Ab initio* calculation method to investigate the slight change of

110 electronic structure in phase transition, which have been also proved by the

111 experimental observation in the Martensitic phase transition of $Ni_2Mn_{1+x}Sn_{1-x}$. Even if

112 the degenerate states of *d* orbits were not changed, the phase transition would also be

113 predicted by the evolution of binding energy in the case of varying composites. We

114 calculated the total energy variation of $Tb_xDy_{1-x}Co_2$ caused by a tetragonal and rhombohedra distortion form the Laves phase. The results for $DyCo_2$ (x=0) were shown in Figure 2a. The tetragonal phase is energetically favorable for $DyCo_2$, as reported previously[14]. As shown in Figure 2a, the Laves phase, which is the parent phase for $Tb_xDy_{1-x}Co_2$ above $T_c$, tends to be unstable against tetragonal and rhombohedral distortion at low temperature. Significantly, $DyCo_2$ with tetragonal phase (AMD symmetry) manifested a favorite structure, when the c-axis was compressed 9.3%. Comparing to the energy evolution of $DyCo_2$ with rhombohedral phase, the tetragonal phase is optimal for $DyCo_2$. Interestingly, the tetragonal phase tends to be unstable against rhombohedral distortion with increasing Tb composition[26]. For $TbCo_2$ (x=1), the minima in the total energy are located at the 86.5° along the rhombohedra distortion (R-3M). Figure 2b shows the total-energy variation of $TbCo_2$ (x=1) caused by the tetragonal or rhombohedra distortion from the Laves phase. The rhombohedra distortion is more energetically favorable than the tetragonal distortion. The theoretical results are consistent with the experimental observation[26, 28], where the $Tb_xDy_{1-x}Co_2$ alloy shows the different symmetry at the Dy-rich side and at the Tb-rich side. Table 1 shows the minima in the total energy of DyCo2 and $TbCo_2$ with the tetragonal and rhombohedra symmetry. Such a phenomenon has perplexed the researchers for a long time, because Co atoms were ligands, occupying the vertex of a truncated tetrahedron and the spin electronic configuration of *f* orbit for Tb or Dy is too complex to be investigated. To clarify this transition and gain further insight into the origin of the MPB observed in $Tb_xDy_{1-x}Co_2$, we have investigated the energy position of peak near $E_F$, which is apparently responsible for the MPB.

Figure 3 shows the total density of states near $E_F$, in the tetragonal phase of $Tb_xDy_{1-x}Co_2$ for x=0, 0.1, 0.2, 0.3, 0.4, 0.5, 0.6, and in the rhombohedral phase of $Tb_xDy_{1-x}Co_2$ for x=0.7, 0.8, 0.9, 1, respectively. A peak was found at about 0.6 eV below $E_F$ in the DOS of $DyCo_2$. The peak in the DOS is predominantly composed of Co 3d sub-orbitals, as pointed out in Figure 1. With an increase in the Tb composition, the peak structure shifts towards $E_F$. The peak shift can be attributed to the hybridization between the Co 3d orbitals and 4f orbitals of excess Tb atoms at Dy sites. It should be noted that the magnetic moment of $Tb_xDy_{1-x}Co_2$ was offered by Co. The substitute of Dy by Tb only changed the electrostatic potentials, which would affect the energy of Co 3d sub-orbitals, rather than the degenerate states. Comparing to the Dy, Tb lacks a 4f electron. Considering the hybridization between the Co 3d orbitals and 4f orbitals of Tb or Dy, the peak in the DOS would shift towards $E_F$. Due to the limitation of the degenerate states of Co 3d orbitals, the peak would not shift through the $E_F$. However, due to the substitute of Dy by Tb, the peak shifts more and more close to the $E_F$. As a 151  result, the phase boundary was located between the Dy-rich side and the Tb-rich side.

152  Our calculation predicted that such a phase boundary should locate around x=0.7 for

153  $Tb_xDy_{1-x}Co_2$, which well agrees the previous experimental observation[14]. Crossing the

154  phase boundary, the peak would shift away from the $E_F$, as the composite of Tb

155  increases. Such a calculation results could be simply understood via studying the

156  relationship between number of valence electrons (NVE) and $E_F$. If the Dy site is

157  substituted by a Tb atom, a smaller NVE would be expected to result a drop of $E_F$. The

158  abnormal shift of the peak in the vicinity of $E_F$ is attributed to the variation of NVE.

159  However, the peak corresponding to Co 3$d$ orbit is prohibited to cross $E_F$. Therefore, the

160  phase transition occurs, when the abnormal shift is reversed.

161  The most interesting feature of the MPB composition is the magneto-responsive

162  properties. Certainly, the MPB would separate two composite dependence free energy

163  evolutions in accordance with symmetry. Figure 4 shows the composition dependence

164  of integrated spin density and Fermi energy. Accompanying the free energy varying, the

165  integrated |spin density| shows a step at the MPB composition, as shown in Figure 4a.

166  The integrated |spin density| denotes the spontaneous magnetization, which is also

167  superiority at Tb-rich side in the experimental observation[14] as shown in the inset of

168  Figure 4a. Figure 4b shows the variation of Fermi level as a function of the composite.

It is significant that a transition occurred at the MPB composition. The MPB separated $Tb_xDy_{1-x}Co_2$ into different symmetries.

In conclusion, we proved a magnetic MPB in a ferromagnetic $Tb_xDy_{1-x}Co_2$ system, which separates two magnetic phase with different crystal symmetries. The MPB leads to a significant variation in magnetic properties and in crystal structure. The calculation reveals that the MPB is a boundary of T and R phase and is a thermodynamically bistable states. The proving of MPB in ferromagnetic system may provide an effective approach for developing highly magnetostrictive materials, and suggests the possibility of MPB in other ferroic systems. The calculation of MPB provides an effective approach to search the MPB composition. It also provides a new insight into the nature of magnetism and phase transition.

**Acknowledgement**

This research was supported by the National Basic Research Program of China (Grant No. 2012CB619401), National Science Foundation of China (Grant Nos. 51222104 and 51071117) and Fundamental Research Funds for Central Universities.

242 **Table list**

243 Table 1. Comparison of total energy between DyCo2 and TbCo$_2$ with the tetragonal and

244 rhombohedra symmetry

245 

|  | Total energy (eV) |  | Favorable structure |
| --- | --- | --- | --- |
|  | Tetragonal | Rhombohedral |  |
| DyCo$_2$ | -48607.9 | -48607.2 | Tetragonal (a=7.12 A, c=6.46A) |
| TbCo$_2$ | -44720.9 | -44721.1 | Rhombohedral (a=7.25 A, α=86.5°) |

247

248

**Captions**

Figure 1. The truncated tetrahedron configuration and the DOS for $DyCo_2$ with Laves phase.

Figure 2. (a). The total energies of $DyCo_2$ with Tetragonal symmetry and Rhombohedral symmetry. (b). The total energies of $TbCo_2$ with Tetragonal symmetry and Rhombohedral symmetry

Figure 3. The total density of states near $E_F$, in the tetragonal phase of $Tb_xDy_{1-x}Co_2$ for x=0, 0.1, 0.2, 0.3, 0.4, 0.5, 0.6, and in the rhombohedral phase of $Tb_xDy_{1-x}Co_2$ for x=0.7, 0.8, 0.9, 1, respectively.

Figure 4. Composition dependence of physical properties in relation with MPB.

261

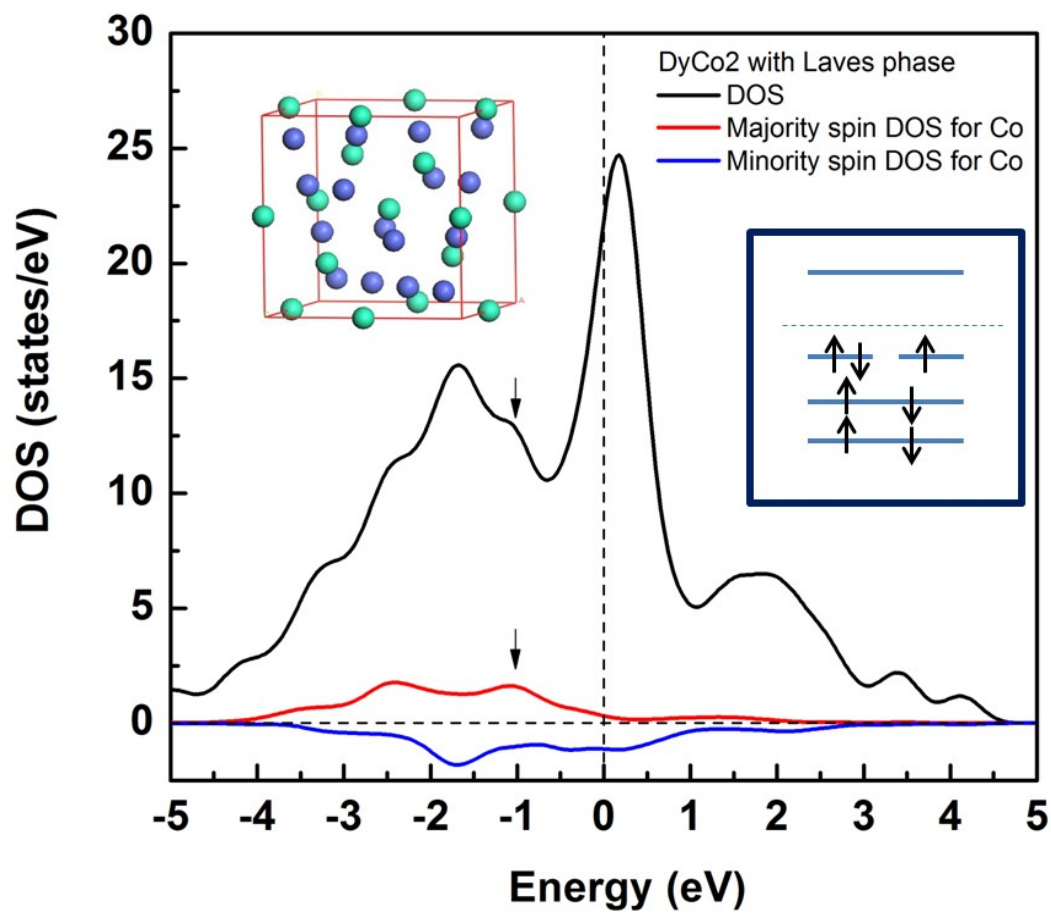

262 Figure 1

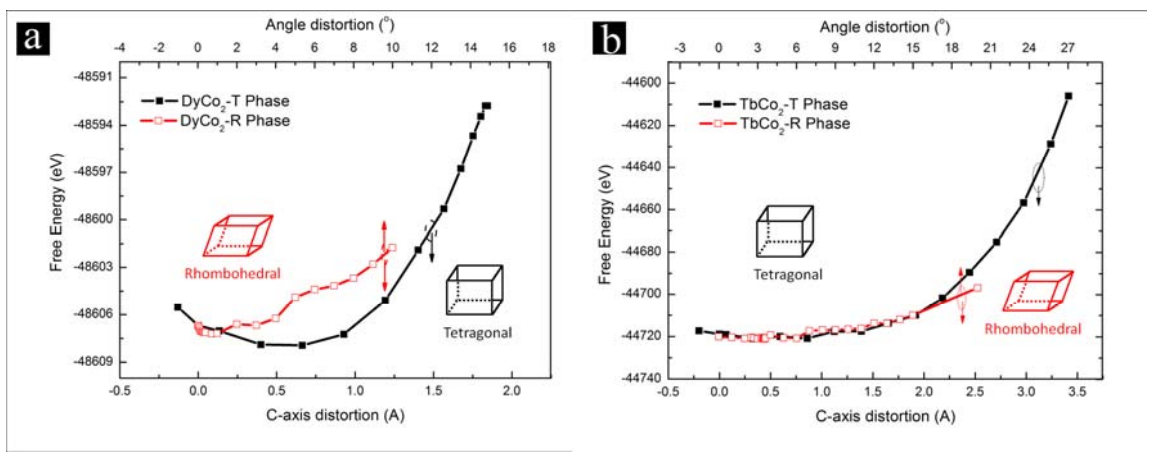

263

264 Figure 2

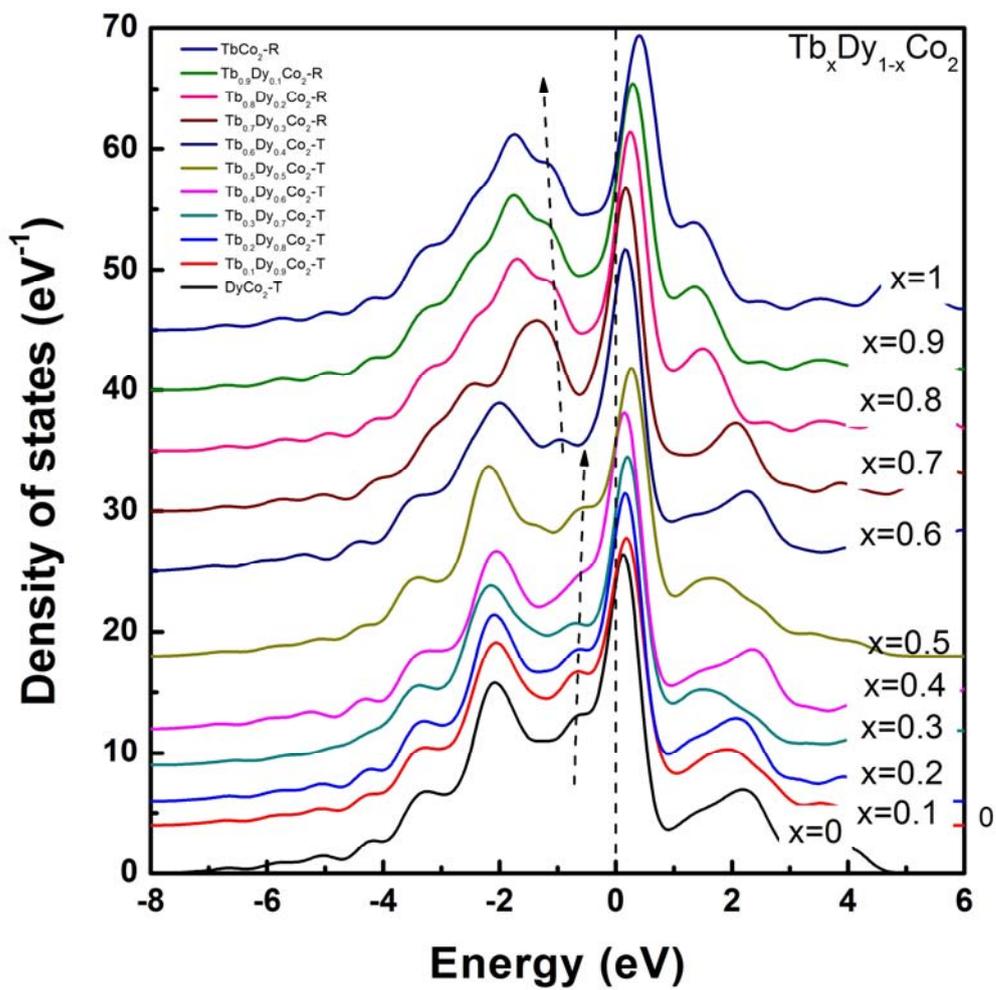

265

266 Figure 3

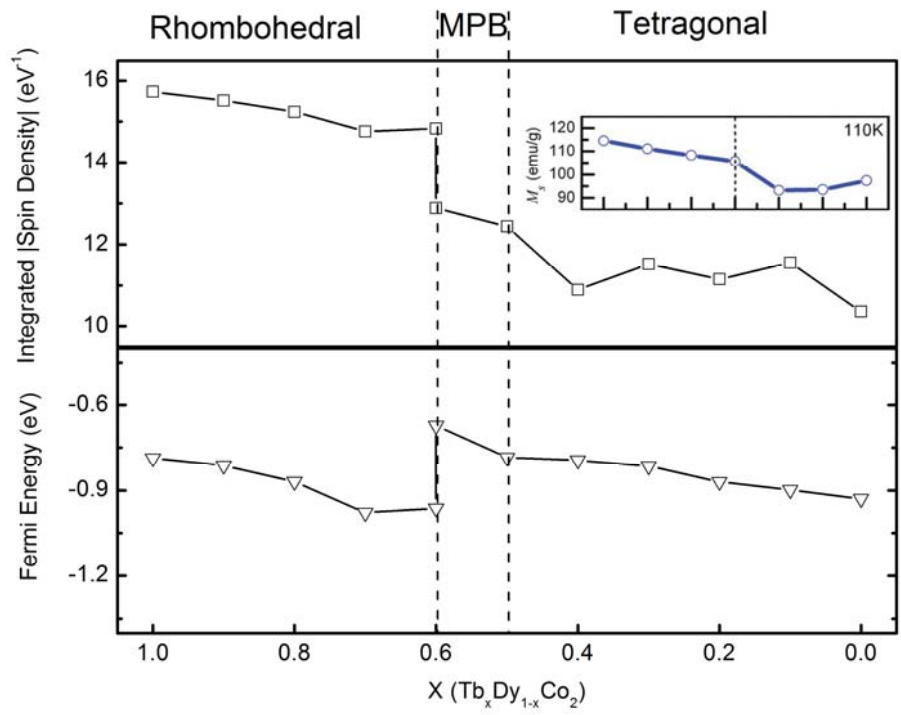

267

268 Figure 4